# Adsorbate-Enhanced Field-Emission from Single-Walled Carbon Nanotubes: A Comparative First-Principles Study

Ehsanur Rahman* and Alireza Nojeh
Department of Electrical and Computer Engineering, University of British Columbia,
Vancouver, BC, V6T 1Z4, Canada
Quantum Matter Institute, University of British Columbia, Vancouver, BC, V6T 1Z4, Canada
*Email: ehsanece@ece.ubc.ca

**Abstract**

We report a comparative ab-initio study of different types of adsorbates and their adsorption mechanism on the field-emission performance of single-walled carbon nanotubes by analyzing electrical properties and transport characteristics and considering the thermal stability of the adsorbed structure. Adsorbates were found to reduce the work function by up to 1.3 eV, enhance tunneling near the carbon nanotube tip, and increase the field-emission current by as much as two orders of magnitude. A significant localization of the electron cloud was also observed near the adsorbates under a high applied electric field .

**I. INTRODUCTION**

Carbon nanotubes (CNTs) are of great interest for electron emission because of their suitable properties such as low turn-on voltage, the strength of the structure, and diversity and flexibility of fabrication processes. There has been extensive research on CNT-based emitters, and a range of applications have been demonstrated [1-3]. The electron emission current depends on several factors among which the work function, $\Phi$, is particularly important due to the exponential dependence of the emission current on this parameter. A great deal of research has been reported so far on reducing the work function of CNTs through the introduction of adsorbates. Improved field-emission from single-walled CNTs (SWCNTs) has also been reported through introducing adsorbates with intrinsic dipole moments such as the water molecule [4] by Grujicic *et al.* or methane [5] by Naieni *et al.* In another study, Yaghoobi *et al.* have investigated the effect of hydrogen adsorption on field-emission from SWCNTs [6]. Lanthanum hexaboride (LaB6) has also been shown experimentally to lower the electron work function of multiwall CNTs (MWCNTs) and improve the field-emission current at low electric fields by Wei *et al.* [7]. In addition, a significant reduction in CNT work function has been reported through adsorption of alkali metals. For example, Kim *et al.* have reported a 1.3-time decrease in turn-on voltage and over one order of magnitude increase in field-emission current using in situ cesium (Cs) treatment of MWCNTs [8]. Wadhawan *et al.* have reported more than two times reduction in turn-on field and six orders of magnitude increase in field-emission current through Cs deposition on SWCNTs [9]. Another study by Shahi *et al.* has reported more than 50 % increase in field-emission current by deposition of cesium iodide (CsI) nanoparticles on CNT field emitters [10]. In addition, theoretical studies conducted by Khazaei *et al.* have shown a 2.5-fold increase in the emission current by introducing one Cs atom over the tip of an armchair SWCNT [11]. Other than Cs, several works have reported work function reduction of CNTs by intercalation of potassium (K). Westover *et al.* have shown that the potassium intercalation of a CNT mat can reduce the work function to approximately 3.1-3.4 eV [12]. In another work, the same group has lowered the carbon nanotube work function to 2-4 eV by doping with potassium atoms using a two-zone vapor method [13]. Our group has also observed a 0.7-1.1 eV reduction in work function during thermionic emission from potassium intercalated MWCNT forests [14]. A challenge is that as temperature increases either due to Joule heating during field-emission or as necessary in thermionic emission, weakly adsorbed species can become desorbed from the CNT surface [14-15]. Therefore, in addition to the work function reduction, the stability of the CNT-adsorbate structure at elevated temperatures is crucial for reliable operation. In this regard, chemisorption, which can provide stronger bonding and also reduce the work function, is desirable. For example, the formation of transition metal carbides near CNT surfaces can be promising [16–18]. In fact, Zhang *et al.* have reported increased emission current, site density and stability when hafnium carbide (Hf-C) bonds were formed on the surface of a CNT film after annealing at $1200^0C$ [19]. The authors attributed this improved performance to a lower work function, chemical inertness, and robustness of the Hf-C bond. A comparative analysis of different CNT-adsorbate structures considering both the thermal stability and the extent of current enhancement with reduced work function is thus needed. Here we attempt a systematic first-principles simulation study of a few experimentally reported CNT-adsorbate systems and present a comparative analysis of these structures considering both their current enhancement capability and thermal stability.

## II. SIMULATION METHODOLOGY

We carried out the simulations on a 90-atom capped single-walled CNT structure which was defined using the Avogadro software package [20]. This carbon structure consists of a 3-unit cell (5,5) metallic armchair CNT which was capped with half a C60 molecule. The capping of the carbon nanotube makes the structure energetically more favourable compared to an open-ended CNT in the presence of an electric field [21]. The dangling bonds on the opposite end of this carbon structure were terminated by hydrogen atoms, which is a common practice for the first-principles quantum mechanical study of CNT structures.

### A. Geometry Optimization

Initial geometry optimization of the bare CNT structure was performed in Avogadro using the Universal Force Field (UFF). The CNT structure was then further relaxed using GAUSSIAN 09 software [22]. The restricted Hartree-Fock (RHF) ab-initio method was used with the 6-31G(d) basis set to perform this geometry optimization. The 6-31G(d) basis set was chosen as it has been reported to represent carbon atoms adequately [23]. To simulate the effect of physisorption on the defined SWCNT tip, an alkali metal atom (Cs or K) was placed near the tip of the capped CNT [fig.1 (a)], and then a geometry optimization step was performed on the entire structure using the same Hartree-Fock (HF) method in order to find the most energetically favourable location of the alkali atom with respect to the CNT tip. The 6-31G(d) basis was again used for all the carbon and hydrogen atoms in the system. However, to reduce computational burden and facilitate convergence of the self-consistent field cycle, an Effective Core Potential (ECP), namely LANL2DZ, was used for alkali atoms. LANL2DZ is considered a shape consistent ECP which usually gives correct bond lengths and structures. It has been developed and used for several heavy elements including Cs to describe the outermost valence electrons [24, 25]. In a similar way, the effect of chemisorption on the CNT tip was simulated by forming a transition metal carbide bond near the pentagon-shaped ring of the CNT tip [fig.1(b)]. Hafnium was chosen for the chemisorption study as the formation of hafnium carbide bonds on CNTs has been observed experimentally in the recent literature [19]. The structure of the chemisorbed CNT was then relaxed in Gaussian software using the same HF theory with the 6-31G (d) basis set for carbon and hydrogen atoms and the LANL2DZ ECP for the hafnium atom. After geometry optimization, the electronic properties of the relaxed structures were simulated under different applied static electric fields. The same HF method with the appropriate basis set for carbon and adsorbate atoms was used for the electric field study on each of these structures.

### B. Electric Field Analysis

To simulate the effect of the electric field, a uniform external electric field was applied along the axis of the carbon nanotube. Due to the high aspect ratio of a typical CNT structure in experiments, the electric field is known to be enhanced significantly near the CNT tip. However, this first-principles simulation study cannot capture the field enhancement effect due to the short length of the nanotube. To consider the field enhancement factor, the electric field analysis in this work was carried out by using an electric field value relevant to tunnelling (several volts per nanometer). The total electric field distribution around the CNT was determined self-consistently by considering both the applied uniform electric field and calculating the charge distribution around the nanotube in response to the applied field. The resulting electrostatic potential distribution around the CNT was extracted for each value of the field. This potential distribution was then used to calculate the field-emission current for both the bare CNT and CNT-adsorbate structures.

### C. Field-emission characterization

To calculate the emission current, the Non-equilibrium Green's function (NEGF) formalism was used which has been reported for the field-emission study of other nanotube structures [26-28]. In this work, using this formalism, the CNT, the adsorbate and the vacuum barrier near the CNT tip were considered as one complete device. The field-emission analysis was then carried out by studying the electronic transport through this composite structure. Electronic transport through a device can be calculated by modelling the device with its Hamiltonian and coupling it to infinitely long contacts [29-30]. The interaction of the contact with the channel is defined by a self-energy term where perturbation of the contact due to the channel is assumed to be on the surface. The device is then connected to the anode and cathode electrodes. The electron transmission probability through the device defined in this manner is then calculated by evaluating the Green's function, which can be defined as

$$G(E) = [(E + i0^+)I - H - \Sigma_1 - \Sigma_2]^{-1} \quad (1)$$

, where $E$ is the electron energy, $H$ is the three-dimensional device Hamiltonian constructed in a real space basis, $i$ is the unit imaginary number, $0^+$ is an infinitesimally small positive number, and $\Sigma_1, \Sigma_2$ are the self-energy terms that account for the coupling of the contacts to the channel. It is worth mentioning that, in the structures studied in this work, the hydrogen termination of the dangling bond at the CNT end creates an undesired local dipole because carbon is more electronegative than hydrogen. Therefore, the electrostatic potential distribution in the structure was extracted from the first-principles HF simulation excluding the hydrogen atoms and used in creating a three-dimensional, real space Hamiltonian. The transmission probability of electrons at different energies was calculated using the Green's function in the Fisher-Lee relation:

$$T = Trace[\Gamma_1 G \Gamma_2 G^\dagger] \quad (2)$$

, where $\Gamma_1 = i[\Sigma_1 - \Sigma_1^\dagger]$ and $\Gamma_2 = i[\Sigma_2 - \Sigma_2^\dagger]$ (3)

The tunnelling currents at different applied electric fields at room temperature were calculated using the Landauer–Buttiker formula:

$$I = \frac{2q}{h} \int T(E)[f_1(E) - f_2(E)] dE \quad (4)$$

, where $f_1(E)$ and $f_2(E)$ are the Fermi functions of the contacts, $T(E)$ is the transmission probability of the electron through the channel at energy E, $q$ is the electron charge, and $h$ is Planck's constant.

### III. RESULTS AND DISCUSSION

Before studying the electronic properties, we briefly discuss the relaxed geometries of different CNT-adsorbate structures. After geometry optimization using the HF method, the average C-C bond length of the bare CNT was found to be 1.43Å, which is consistent with the values reported in the literature [31-32]. For the CNT-adsorbate structures, the position of the adsorbate near the CNT tip depends on the adsorption mechanism and on the type of adsorbate atom. For physisorption of alkali metal atoms, the stability of the physisorbed structures was verified by initially placing the adsorbate at different positions near the CNT tip, which resulted in almost identical relaxed geometries. For chemisorption of the Hf atom on the CNT tip, the average Hf-C bond length after geometry optimization was found to be 2.13Å, which is close to the values reported in other first-principles studies [33-34]. The structural stability of the Hf-chemisorbed CNT was further verified through geometry relaxation using density functional theory (DFT) with the Perdew, Burke, and Ernzerhof (PBE) exchange-correlation functional.

The strength of the adsorption process can be an indicator of the thermal stability of the CNT-adsorbate structure. To compare the thermal stability, we calculate the binding energy of these different adsorbates to the CNT tip using the following equation:

$$E_{Adsorption} = E_{CNT} + E_{Adsorbate} - E_{CNT+ Adsorbate} \quad (5)$$

Using the above equation, the binding energies of the physisorbed Cs and K atoms to the CNT were found to be 0.1363eV and 0.1350 eV, respectively. Similarly, the binding energy of the Hf-C bond formed during the chemisorption process was calculated to be 1.12 eV which is much higher than the physisorption process. These binding energy values further confirm the better thermal stability of the chemisorbed structure compared to the physisorbed ones.

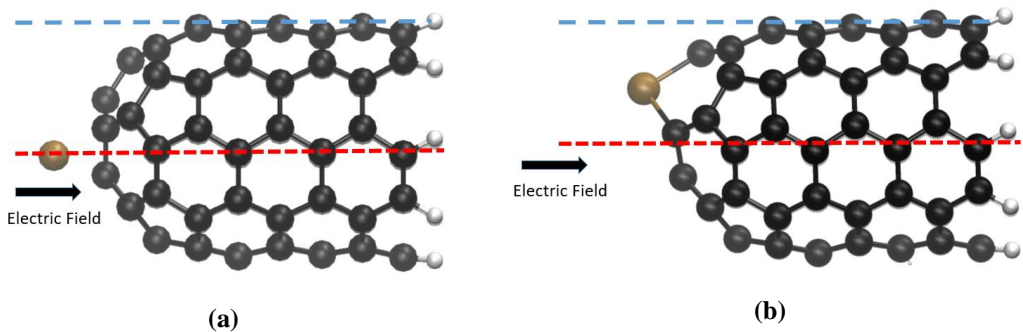

Fig.1 (a) Physisorption and (b) Chemisorption of an adsorbate on the tip of a SWCNT. The dashed blue and red lines indicate the CNT wall and axis, along both of which the electrostatic potential energy profile was calculated.

To study field-emission from the CNT-adsorbate system, we examine the effect of these adsorbates on the electrostatic potential energy along the direction of the applied electric field. In field-emission, reducing the width and height of the potential barrier which the electrons must overcome to reach vacuum is crucial for improving the efficiency of the emission process. This can be investigated through the study of the electrostatic potential energy along the different planes of the pristine CNT and CNT-adsorbate system. A detailed comparative analysis of the effect of the adsorption mechanism and adsorbate type on field-emission can be carried out based on fig. 2. Figs. 2(a)-(d) shows the electrostatic potential energy along the wall as well as the axis for both the pristine CNT and CNT-adsorbate structures without and with an electric field. In the potential energy profile along the axis [Figs. 2(b), (d)], a potential barrier is observed near the tip of the CNT. As the electric field increases, this barrier is lowered and electrons can tunnel out more easily. The extent of this barrier lowering under the application of an electric field depends on the applied electric field, the presence of the adsorbate, the adsorption mechanism, and the adsorbate type. In the case of CNT-alkali metal adsorbate structures, a second potential well appears near the adsorbate atom as can be seen from figs. 2(b) and (d). Electron emission from these systems occurs due to tunnelling from this potential well. This point will be discussed further by studying the spatial distribution of charge and the highest occupied molecular orbital (HOMO) later in this work. On the potential profile along the CNT wall [Fig. 2(c)], there is also a noticeable reduction in barrier width and height near the CNT tip of the CNT-adsorbate structures under an applied electric field. The extent of this lowering is different for different CNT-adsorbate structures. Moreover, without an applied electric field, the HOMO energy level is shifted closer to the vacuum level due to adsorbates.

Since the CNT adsorbate systems have significantly lowered the height and width of the potential barrier for the tunnelling electrons compared to the pristine CNT case, we now turn to a quantitative comparison of this potential barrier reduction for different CNT-adsorbate and the pristine CNT structure. Fig. 2(c) shows that, at an applied electric field of 0.5V/Å, the width and height of the potential barrier along the CNT wall that the HOMO electrons would have to tunnel through to escape into vacuum are 1.2 Å, 5.5 Å, 8.5Å, and 12.15 Å and 0.24 eV, 3.87 eV, 3.21 eV, and 3.82 eV for the CNT-Cs, CNT-K, CNT-Hf and bare CNT structures, respectively. On the other hand, as shown in fig. 2(d), in the presence of adsorbates, the potential barrier along the axis was already overcome for the HOMO electrons near the CNT tip under an applied electric field of 0.5V/Å. From these comparisons, it can be said that these adsorbates can enhance the electron tunnelling probability from the CNT tip under an applied electric field. However, the alkali metal adsorbates create a more favourable tunnelling state for field-emission compared to the chemisorbed Hf.

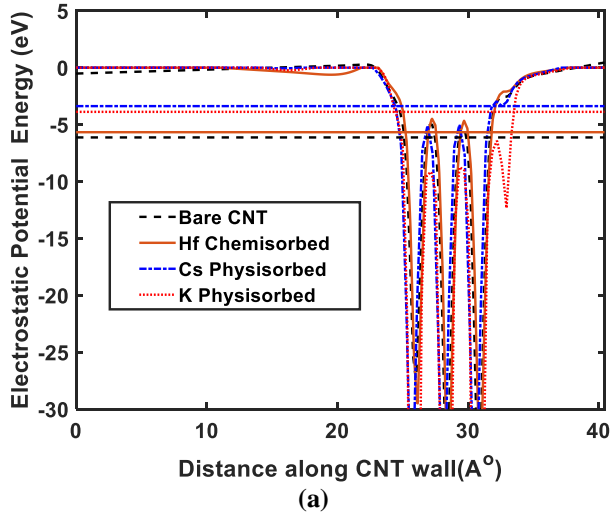

(a)

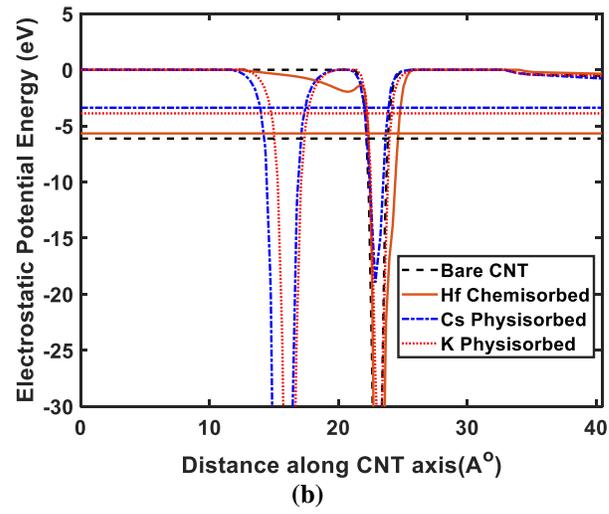

(b)

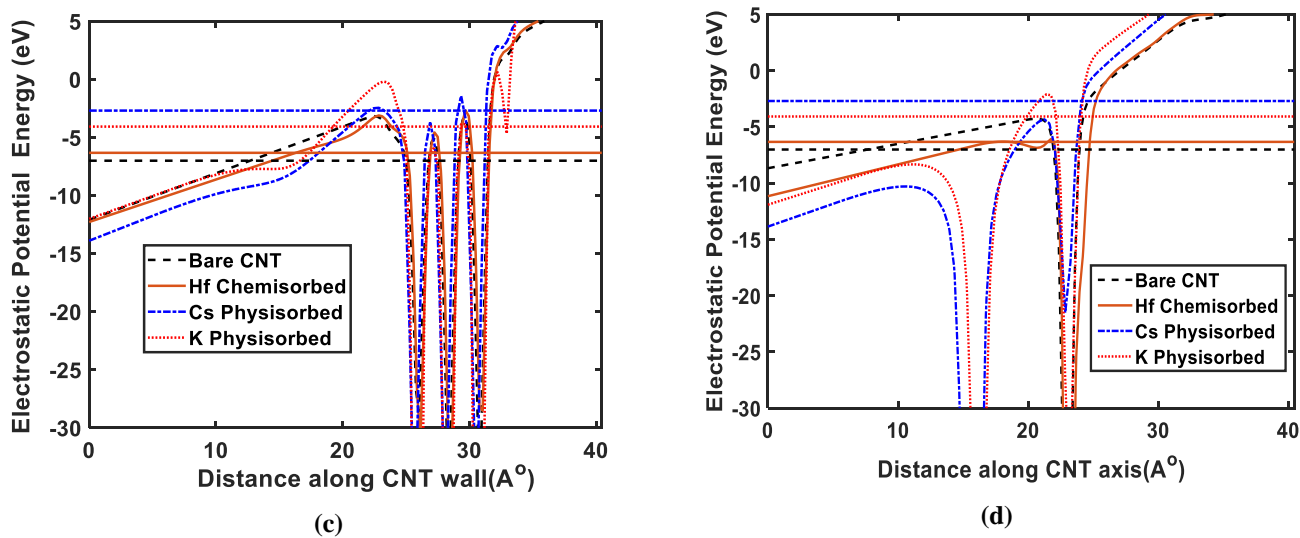

Fig. 2. Effect of physisorption and chemisorption on electrostatic potential energy distribution with no electric field [(a)-(b)] and when a 0.5V/Å electric field is applied [(c)-(d)]. The horizontal lines represent the HOMO energy levels for bare CNT and different CNT-adsorbate structures.

For further analysis, It would be instructive to investigate the effect of these adsorbates on charge distribution near the CNT tip. Fig. 3 shows the Mulliken charge population for the bare CNT as well as different CNT-adsorbate systems under an applied electric field. For a pristine CNT, under an electric field of 1V/ Å [Fig. 3(b)], five carbon atoms on the pentagon-shaped ring at the tip have the highest charge of 0.067 electrons. In case of the CNT-adsorbate structure, an accumulation of charges occurs near the CNT tip and the adsorbate. (It is worth mentioning that the stability of the physisorbed structure under this field strength was verified by performing a series of calculations of the total energy of the adsorbed structure as a function of the position of the adsorbate with respect to the nanotube tip. The observed stability of the physisorped structure is due to the dipole moment induced between the adsorbate and the SWCNT in the presence of an electric field). However, there is a significant difference in the magnitude and polarity of charge accumulation under different electric fields for the physisorbed and chemisorbed adsorbates. For example, in the CNT-Cs structure, under an electric field of 0.5 V/ Å (not shown), the charge on the Cs atom is 1.07 electrons, which is much higher than that of the C atoms of the pristine CNT tip even at the highest electric field of 2V/ Å(not shown). Another interesting observation is that, as the electric field increases, the C atoms at the tip gradually become depleted of electrons as observed by the change in charge polarity for C atoms in the CNT-Cs structure. A similar charge transfer was found for the CNT-K adsorbate system although the amount of charge accumulated on the K adsorbate at the applied electric field was slightly lower than that of Cs. In terms of electron emission sites, the emission would efficiently take place from the adsorbate atom in case of the CNT-Cs and CNT-K structure, and from the five C atoms at the tip of the pristine CNT. Consequently, one expects a considerable localization in the electron emission site due to the physisorption of one alkali metal atom. On the other hand, for the chemisorption of Hf, there is considerable charge separation even without an electric field [Fig. 3(g)]. Due to the higher electronegativity of C atoms compared to Hf, they accumulate a negative charge of about 0.43 electron each. (The positive charge on the Hf atom is equal to the sum of the negative charges accumulated on each carbon to which it is bonded.) When an electric field is applied, electrons start accumulating on the Hf adsorbate but to a lower extent compared to the physisorbed alkali metals at each electric field. In fact, the accumulated negative charge on the Hf atom exceeds the electronic charges on neighbouring C atoms to which it is bonded only when the applied electric field is 1.5V/ Å (not shown). It should also be noted that, unlike in physisorption of alkali metals, the Hf and constituent carbon atoms forming the Hf-C bond will all work as electron emission sites [Fig. 3(h)] at the high electric field, leading to less localization of emission compared to the former. The electrostatic potential energy profile around the different CNT structures, resulting from the charge distribution on the atoms without and with an applied electric field, is shown in fig. S1 of the supplementary document.

To provide more insight, it is worth investigating the spatial extent of the electron emission sites of the bare CNT and different CNT-adsorbate structures under study. Electron emission sites on CNTs have been imaged using field-emission microscopy [35-36]. The strength of the electric field near the emitter surface in typical field-emission experiments seems to be on the order of several volts per nanometer [37]. We examined the spatial extent of the HOMO cloud for both the bare CNT and different CNT-adsorbate structures for a wide range of applied electric field (up to 2V/ Å) to follow the gradual evolution of the electron emission site in these structures. It is observed that the HOMO cloud starts accumulating near the tip for the bare CNT as the applied electric field value approaches 1V/Å (the gradual evolution of the HOMO cloud for the pristine CNT at different electric field strengths is shown in fig. S2 of the supplementary document). An accumulation of the HOMO cloud near the adsorbate is also observed for the different CNT-adsorbate structures, although at a lower electric field strength compared to the bare CNT. Fig. 4 shows the HOMO cloud distribution for the pristine CNT and different CNT-adsorbate systems at the electric field of 1V/ Å. For the pristine CNT, the HOMO cloud covers mainly the carbon atoms near the tip. In the Hf chemisorbed structure, the Hf and the associated C atoms forming the Hf-C bond are strongly covered by the HOMO cloud. However, in the case of physisorption of alkali metals near the CNT tip, the HOMO cloud can be found only around the adsorbate for both Cs and K atoms suggesting that only the adsorbate atom effectively takes part in electron emission at a high electric field.

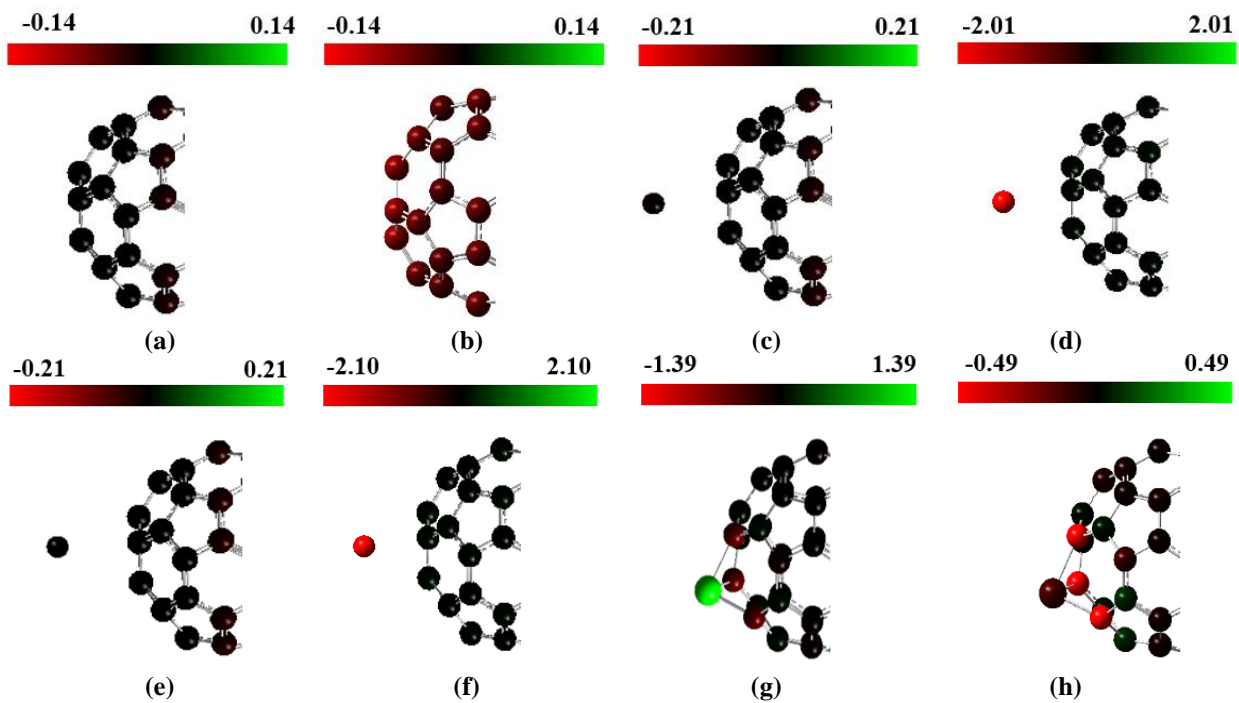

Fig. 3. Mulliken charge population in a bare CNT [(a)–(b)], CNT-K structure [(c)–(d)], CNT-Cs structure [(e)–(f)], and CNT-Hf structure [(g)–(h)] under no electric field [(a), (c),(e),(g)] and an electric field strength of 1 V/ Å [(b),(d),(f),(h)].

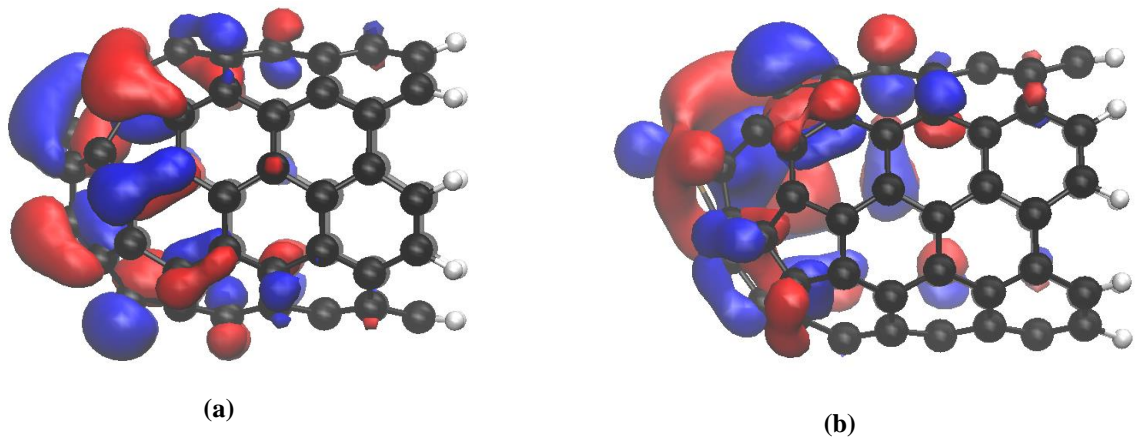

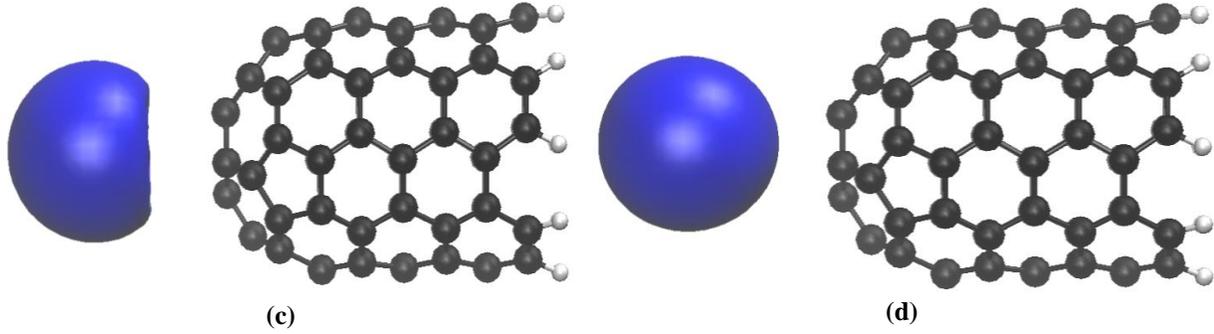

**(c)**          **(d)**

Fig. 4. Spatial distribution of HOMO cloud for the (a) pristine CNT (b) CNT-Hf structure (c) CNT-Cs structure and (d) CNT-K structure under an electric field strength of 1 V/ Å. The surfaces shown correspond to a contour value of 0.02.

Considering the effects of these different adsorbates on the electrostatic potential barrier and electron emission sites, it is worth quantifying the combined result of these effects on field-emission performance by calculating the current for these different structures. Fig. 5 shows the emission current versus electric field for the pristine CNT and various CNT-adsorbate systems. An important observation from this figure is the current saturation behaviour at high electric fields for both pristine and CNT-adsorbate structures. Such current saturation has also been observed experimentally in other CNT emitters [9,38-41]. The theoretical origins of the current saturation behaviour have been discussed in [27], based on the evolution of the potential barrier near the nanotube tip as a function of the applied electric field. As seen from Fig. 5, the current is significantly enhanced for the CNT-adsorbate system, as expected from the discussions above. The highest current was observed for the CNT-Cs adsorbate structure. (We emphasize that, due to the simple nature of the transport model, the calculated current values are useful for qualitative comparisons, rather than as a strict quantitative prediction.)

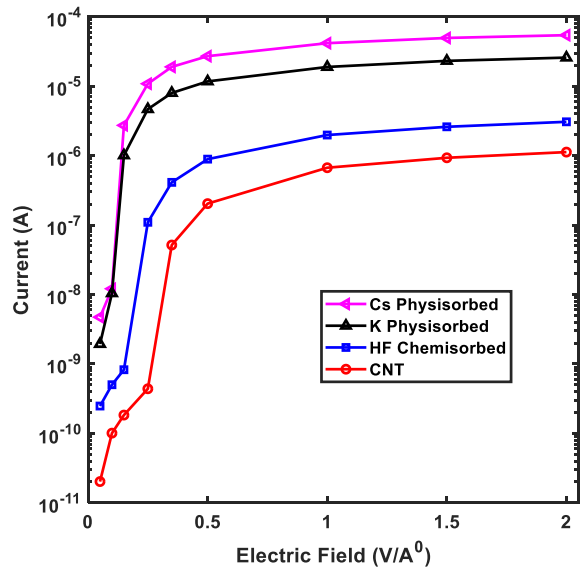

Fig. 5. Emission current as a function of applied electric field for the bare CNT and different CNT-adsorbate structures.

In the case of alkali metal physisorption, we observed about two orders of magnitude increase in current for the CNT-K and CNT-Cs structure compared to bare CNT at an electric field value of 0.5V/Å. On the other hand, for chemisorption of Hf, an increase in current was less than an order of magnitude even at the highest electric field of 2V/Å. Nonetheless, the thermal stability of the chemisorbed structure could make the system more reliable and stable as an electron emitter, especially if the emitter temperature is increased due to field-emission [42-43].

Considering the current enhancement due to the different adsorbates, it would be interesting to investigate the effect of these adsorbates on the work function of the CNT. It is naturally tempting to fit the current-voltage characteristics shown in Fig. 5 to the Fowler-Nordheim (FN) equation, as is common practice in the literature. However, extreme caution is required since the applicability of the FN theory, at least in its original form, to nanoscale emitters is highly questionable—an issue that is too often overlooked. For a thorough study of this issue and the relevant developments and improvements, we refer the reader to the extensive body of literature by Forbes and colleagues [44], Jensen [45] and others. Nonetheless, for a very rough and qualitative comparison, based on an FN fit, we estimate the work function for the CNT-Cs, CNT-K and CNT-Hf structures to be 3.2 eV, 3.5 eV and 4.1 eV, respectively, assuming the work function of the bare CNT to be 4.5 eV. The work function of the CNT emitter thus appears to be significantly reduced by the physisorption of alkali metals compared to the Hf chemisorbed structure. However, the strength of the chemisorbed structure could yield a more reliable and stable performance.

## IV. CONCLUSIONS

In summary, we investigated the effect of different types of adsorbates and their adsorption mechanism on field-emission behaviour from a SWCNT by studying the relevant electrostatic and transport characteristics. The adsorbates increased the field-emission current and reduced the emitter work function to different extents. The increase in field-emission in the presence of adsorbates was found to be due to the adsorbate enhanced tunnelling . This is supported by the potential barrier lowering near the physisorbed or chemisorbed CNT tip under an applied electric field. The adsorbates also change the shape of the electron emission site and localize the emission site further near the CNT tip region at the high electric field as evident from both the charge population and distribution of the HOMO wavefunction in these CNT-adsorbate structures. The increase in field-emission current and reduction in emitter work function were significantly higher for the physisorption of alkali metals compared to the chemisorption of a transition metal adsorbate. However, the thermal stability of the chemisorption process would make the structure more robust to high-temperature operation compared to the physisorbed structure, which has been also reported experimentally. Therefore, this study sheds light on the trade-offs between thermal stability and enhanced field-emission properties due to different adsorbates.

**Acknowledgement:**

We acknowledge financial support from the Natural Sciences and Engineering Research Council of Canada (RGPIN-2017-04608, RGPAS-2017-507958, SPG-P 478867). This research was conducted, in part, by funding from the Canada First Research Excellence Fund, Quantum Materials and Future Technologies Program. Ehsanur Rahman acknowledges financial support from the University of British Columbia for the International Doctoral Fellowship. This research was enabled in part by support provided by WestGrid (www.westgrid.ca), SHARCNET (www.sharcnet.ca) and Compute Canada (www.computecanada.ca).

10